\documentclass[12pt]{article}

\usepackage{a4wide}

\usepackage{amsmath,amssymb,amsfonts} 

\usepackage{extarrows}

\usepackage{mathrsfs,latexsym}

\newcommand{\Ett}{{\tt E}}

\newcommand{\II}{{\boldsymbol{1}}}

\newcommand{\CC}{{\mathbb C}}

\newcommand{\RR}{{\mathbb R}}

\newcommand{\NN}{{\mathbb N}}

\newcommand{\CoinfM}{C_0^\infty(M)}

\newcommand{\CoinX}[1]{C_0^\infty({#1})}

\newtheorem{Thm}{Theorem}[section]

\newtheorem{Def}[Thm]{Definition}

\newtheorem{Prop}[Thm]{Proposition}
\newtheorem{Cor}[Thm]{Corollary}

\numberwithin{equation}{section}

\newcommand{\HH}{{\mathscr H}}

\newcommand{\Ac}{{\mathcal A}}
\newcommand{\Bc}{{\mathcal B}}

\newcommand{\Dc}{{\mathcal D}}

\newcommand{\supp}{{\rm supp}\,}

\newcommand{\dvol}{d\textrm{vol}}

\newcommand{\ip}[2]{{\langle #1\mid #2\rangle}}

\newcommand{\ux}{\underline{x}}

\newcommand{\WF}{{\rm WF}\,}

\newcommand{\Hc}{{\mathcal{H}}}

\newcommand{\Nc}{{\mathcal{N}}}

\DeclareMathOperator{\sinc}{sinc}

\begin{document}

%

\title{The Necessity of the Hadamard Condition}
\author{Christopher J Fewster${}^{(1)}$\thanks{\tt chris.fewster@york.ac.uk}~
and Rainer Verch${}^{(2)}$\thanks{\tt verch@itp.uni-leipzig.de}\\[12pt]
\small ${}^{(1)}$ Department of Mathematics,
                 University of York,
                 Heslington,
                 York YO10 5DD, U.K. \\
\small ${}^{(2)}$ Institut f\"ur Theoretische Physik,
Universit\"at Leipzig,
04009 Leipzig, Germany}
\date{\small \today}
\maketitle

\begin{abstract}
Hadamard states are generally considered as the physical states for linear quantized
fields on curved spacetimes, for several good reasons. Here, we provide a new
motivation for the Hadamard condition: for ``ultrastatic slab spacetimes'' with compact
Cauchy surface, we show that the Wick squares of 
all time derivatives of the quantized Klein-Gordon field have finite fluctuations only if
the Wick-ordering is defined with respect to a Hadamard state. 
This provides a converse to an important result of Brunetti and Fredenhagen. The recently proposed ``S-J (Sorkin-Johnston) states'' are shown, generically, to give
infinite fluctuations for the Wick square of the time derivative of the field, further limiting their utility as reasonable states.  
Motivated by the S-J construction, we also study the general question of extending states
that are pure (or given by density matrices relative to a pure state) on a double-cone region of Minkowski space. We prove a result for general quantum field theories showing that 
such states cannot be extended to any larger double-cone without encountering singular behaviour
at the spacelike boundary of the inner region. In the context of the Klein-Gordon field this shows that even if
an S-J state is Hadamard within the double cone, this must fail at the boundary. 

\end{abstract}

\section{Introduction}

One of the fundamental difficulties in the theory of quantum fields
on curved spacetimes is that generic spacetimes possess no symmetries 
that could serve to distinguish a preferred vacuum state. Instead, 
for linear fields, experience has led to the delineation of the class
of Hadamard states \cite{KayWald:1991} whose short-distance structure
approximates that of states with finite energy density in Minkowski
space, motivated by the equivalence principle. The two-point function
of a Hadamard state $\omega$ of a Klein--Gordon field is required to take the form 
$$ W_\omega(x,y) = \lim_{\epsilon \to 0^+}\,\frac{U(x,y)}{\sigma_\epsilon(x,y)}
 + V(x,y)\ln(\sigma_\epsilon(x,y)) + H_\omega(x,y) $$
upon writing the distribution $W_\omega$ formally as a function of spacetime
points $x$ and $y$. Here, $\sigma_\epsilon(x,y)$ denotes the squared geodesic
distance between $x$ and $y$, together with a suitable regularization, $U$ and 
$V$ are $C^\infty$ functions determined by spacetime metric and Klein-Gordon
equation, while $H_\omega$ is a $C^\infty$ function containing the state-dependence.
(For full details of the definition, see~\cite{KayWald:1991}). It is worth noting that for
static spacetimes, ground states or thermal equilibrium states of the quantized Klein-Gordon 
field are quasifree Hadamard states~\cite{SahlmannVerch_passivity:2000}. Hadamard states
play an important role in computations, for they permit the 
evaluation of Wick polynomials, including the stress-energy tensor, and time-ordered products \cite{BrFr2000}. An elegant reformulation of the Hadamard
condition using microlocal analysis, due to Radzikowski~\cite{Radzikowski_ulocal1996}, has provided sufficiently good
control on the class of Hadamard states that many results
can be proved for general Hadamard states on general spacetime
backgrounds. Examples include Quantum Energy Inequalities \cite{Fews00,FewsterVerch_dirac}, no-go results concerning chronology-violating spacetimes \cite{KRW:1997} and existence results for the semi-classical
Friedmann equations \cite{Pinamonti:2011}. Hadamard states also
play an important role in understanding why black holes display
the Hawking temperature \cite{KayWald:1991}. It is also worth
emphasizing that the class of Hadamard states is defined
in a fully local and covariant way~\cite{BrFrVe03}. 

Nonetheless, one might wonder whether a criterion based on
ultrashort distance behaviour is physically well-founded, especially 
in view of the widespread expectation that the continuum model
of spacetime should break down at very small scales. An alternative
approach might be to start with a model of fundamental spacetime
structure and extrapolate macroscopic state selection criteria from that. 
Just such a proposal was made recently by Afshordi, Aslanbeigi and Sorkin \cite{AAS} based on ideas developed in the causal set programme~\cite{Johnston:2009,Sorkin:2011}, and giving a novel
proposal for assigning a distinguished ``S-J state'' of the Klein--Gordon field to suitable spacetimes or spacetime regions. Further results
appear in \cite{SorkinGmbH,BrumFredenhagen:2013,AslanbBuck:deSit2013}. The idea is the following: consider a globally
hyperbolic spacetime $M$ with metric $g$ such that $\Ett$, the difference of the advanced and retarded Green functions, extends to a bounded operator 
on $L^2(M,d {\rm vol}_{M})$. Then the operator $A = i \Ett$ is
symmetric and [its closure] possesses the polar decomposition
$A = U |A|$. One can form the positive part of $A$, given by 
$A^+ = (1/2)(|A| + A)$, and define a two-point function
\begin{align} \label{eq:DefSJ}
W_{SJ}(f,h) & = \langle \overline{f}, A^+ h \rangle \,, \quad f,h \in \CoinfM\,,
\end{align}
where the bar denotes complex conjugation and $\langle f,h\rangle = \int_M \bar{f}h\,d{\rm vol}_M$ denotes
the $L^2$-scalar product; this defines the 
two-point function of the S-J state~\cite{AAS}. 

In \cite{CF-RV-ultraSJ}, we analyzed the the S-J prescription and showed
that $W_{SJ}$ is indeed the 2-point function of a pure quasifree state on the algebra of smeared Klein--Gordon fields on $M$ whenever
the prescription is defined, and that this certainly holds whenever $M$ can be isometrically embedded as a globally hyperbolic and relatively compact sub-spacetime inside a larger globally hyperbolic spacetime. 
However, we also showed that the S-J states suffer from serious
problems. In particular, they generically fail
to be Hadamard when computed for ultrastatic `slab' spacetimes with
compact spatial section; moreover, they fail to be defined in a locally covariant way. This brings into sharp focus the question of how
seriously the Hadamard condition should be taken in the context of
models inspired by discontinuous fundamental spacetime structure. 

In the present paper,  we aim to bring further clarity to a number 
of issues surrounding the desirability of the Hadamard condition and
the problems that must be faced if one aims to define a state 
on a local region of spacetime. We emphasize that our focus is
much broader than  S-J states, although
they have provided a motivation for some of our results. 

First, we consider general pure quasifree states of the Klein--Gordon field on general ultrastatic slab spacetimes 
with compact spatial section. Any such state $\omega$ can
be used as the basis of a normal ordering prescription and the
construction of Wick powers. Of course, all Wick powers (other than the zero'th order) 
will necessarily have vanishing expectation value in the
basic state $\omega$; this would apply to the stress-energy tensor as well.  
However, in order to use $\omega$ as the basis for a fully fledged
perturbative construction of a self-interacting theory, or for an
analysis of the semiclassical Einstein equations, one would need
the Wick polynomials to have finite (and preferably small) fluctuations
in the state $\omega$. Our main result of Section~\ref{sect:fluctuations}  is that
$\omega$ has finite fluctuations for all Wick polynomials only if 
it is Hadamard (the converse statement is established, for all
spacetimes, in \cite{BrFr2000}). In particular, 
we show that the Wick square of the first time-derivative of the 
field has infinite fluctuations in the S-J state. However, the
main force of our result is to provide a motivation for the
Hadamard condition that is less tightly bound to the ultrashort
distance structure. 

While these results cast serious doubt on the utility of S-J states
defined on spatially compact ultrastatic slab spacetimes, strengthening
the findings of our previous paper~\cite{CF-RV-ultraSJ}, we have not
yet addressed the properties of S-J states defined on bounded
regions of Minkowski space. Our second result concerns this question, 
indeed, we will address it in the model-independent context of
operator-algebraic quantum field theory. In this setting, we will show
that no state that is pure (or even normal to a pure state, i.e.\
arising as a density matrix state in the GNS Hilbert space representation
of a pure state)
on the algebra of observables of a double cone region can
extend to the observable algebra of a strictly larger double cone region 
without developing pathologies at the spacelike boundary of
the smaller double cone: the extended state necessarily fails 
to admit a stable short-distance scaling limit at any point of this boundary.
To borrow a term from recent controversial discussions of black hole evaporation~\cite{AMPS_firewall:2013}, one might say that the
state on the inner region is protected by a firewall (or, `energetic
curtain'~\cite{Braunstein:curtain}).  
In the particular example of the free Klein--Gordon field, the S-J state
of any double cone is pure~\cite{CF-RV-ultraSJ} and our result
entails that any extension to a larger double cone must fail to be
Hadamard at all points of the spacelike boundary, because the 
Hadamard condition implies the existence of a stable scaling limit. 
The argument relies on the type of local von Neumann algebras of the 
double cones. It is well known that under standard assumptions the 
local von Neumann algebras in quantum field theory are of type III, 
while the von Neumann algebra induced by a pure state is of type I. 
In our case, we draw on results taken 
from~\cite{Fredenhagen:1985,BaumWolle:1992} (see also \cite{BuchVer:1995}) establishing that the 
type III property follows from the existence of suitable stable scaling limits which are expected to correspond to the theory
admitting an ultra-violet renormalization group fixed point. Our result
is consistent with the computation performed in~\cite{SorkinGmbH},
where a two-dimensional diamond region was considered for the
massless free field -- near the spacelike boundary the two-point function is approximated by that of the ground 
state of the field in the presence of a mirror at the boundary, so extension through the boundary will not be possible while maintaining the Hadamard form. 

The result just described indicates that S-J states defined on 
extendible spacetime regions have undesirable features and 
the case of double cones seems to suggest that this has to do with 
the `sharp cut-off' at the boundary of the spacetime region on which
they are defined. It has been suggested to us by various people
(we thank Jorma Louko and Rafael Sorkin in particular) that perhaps
the Hadamard condition could be restored for a modified version
of the S-J state prescription in which the `sharp boundary' is softened. 
We study two simple such variants gained by averaging procedures,
in the ultrastatic slab situation, and show explicitly that neither leads
to Hadamard states. Very recently, however, Brum and Fredenhagen \cite{BrumFredenhagen:2013}
have suggested an altered definition of S-J states on certain
globally hyperbolic slab spacetimes which
corresponds to a different `softening-the-boundary' procedure, and their proposal
leads to Hadamard states. This indeed indicates that the failure of the original definition
of S-J states to produce Hadamard states stems from the presence of the `sharp boundary'
in timelike direction in the case of ultrastatic slab spacetimes considered in~\cite{CF-RV-ultraSJ}. However, the altered definition
of \cite{BrumFredenhagen:2013} involves a dependence on a smoothing function as an extra parameter 
--- which is in the very nature of smoothing procedures --- and this is at variance with
the original idea that a unique S-J state can be assigned to any spacetime without the need of specifying
other parameters or possible choices \cite{AAS}. Moreover, it is emphasized in~\cite{BrumFredenhagen:2013} that the modified states are unlikely to carry the interpretation of vacuum states. Nevertheless, the proposal of \cite{BrumFredenhagen:2013} does lead
to a new method of constructing Hadamard states for certain static or expanding slab spacetimes, and it would be of interest to see if that could be extended to a larger class of spacetimes. 

As mentioned in \cite{BrumFredenhagen:2013}, the definition of S-J states for the quantized
scalar Klein-Gordon field seems to have had a forerunner for the case of the quantized Dirac field
in the form of the `Fermionic projector' proposed some time ago by Felix Finster \cite{Finster-DiracSea},
see also the recent paper \cite{FinsterReintjes:2013}.
(We would like to thank Felix Finster for some comments on that point.) 

The organization of the present work is as follows. In Section 2, we establish that the requirement of
finite variance for
the Wick squares of time-derivatives of any order of the quantized Klein-Gordon field on an
ultrastatic slab spacetime with compact Cauchy surface implies that the states with respect
to which the Wick-ordering is defined must be Hadamard. We also show that S-J states do not
comply with this requirement as Wick squares of the time-derivative of the field have
infinite fluctuations for S-J states.

In Section 3, we show for general quantum field theories on
Minkowski spacetime that a state on the algebra of observables of double cone whose induced 
GNS Hilbert space representation is of type I cannot be extended to a state on the observable
algebra of a larger spacetime region unless the state fails to have a stable short-distance
scaling limit at every point of the double cone's (spacelike) boundary. Such an assertion may be
known to some experts, but would surely be hard to trace in this form in the literature. We
then apply this result to show for the case of the quantized Klein-Gordon field on Minkowski
spacetime that S-J states defined for double cones cannot be extended to states on observable
algebras of larger spacetime regions which are normal to Hadamard states. (The result generalizes,
as we shall indicate, also to certain curved spacetime situations.)

In Section 4, we present two simple averaging procedures for S-J states defined for the quantized Klein-Gordon
field on ultrastatic slab spacetimes, and we show that these procedures do not produce Hadamard states.
We also show that our previous results in \cite{CF-RV-ultraSJ} on S-J states failing to be Hadamard
for ultrastatic slabs can be strengthened by invoking a local-to-global result on the Hadamard property
of two-point functions of states due to Radzikowski \cite{Radzikowski:loc-to-glob}. 
In particular, if the Cauchy-surface
of the ultrastatic slab spacetime is acted on by a transitive group of isometries, then the 
corresponding S-J state does not only fail to be Hadamard on the ultrastatic slab spacetime as a whole
(i.e., `somewhere'),
but it fails to be Hadamard everywhere. 

The article is concluded by a discussion in Sec.\ 5.

\section{Fluctuations in Wick polynomials and Hadamard property}\label{sect:fluctuations}

In this section, we investigate how the requirement that 
certain Wick polynomials have finite fluctuations enforces 
the Hadamard condition. In passing, 
the S-J states on ultrastatic slabs are shown to produce infinite fluctuations for smearings of ${:}\dot{\phi}^2{:}$ 
except for at most a set of measure zero in $\tau$. Throughout the section, 
we will say that a field has finite fluctuation in a state if all of its smearings with test functions have 
finite fluctuations in that state.

We consider the massive scalar field $(\Box+m^2)\phi=0$, $m>0$, on $d$-dimensional ultrastatic slab spacetimes of the form $M=(-\tau,\tau)\times\Sigma$
with metric $ds^2 = dt^2 - h_{ij} dx^i dx^j$, where $\Sigma$
is a compact manifold with smooth Riemannian metric $h$ and $\tau\in (0,\infty]$. Then there exists a complete orthonormal basis
$\psi_j$ ($j\in J$) for $L^2(\Sigma)$ (with the volume measure induced by $h$),
comprising eigenfunctions of $K=  -\triangle + m^2$, where $\triangle$ is the Laplacian on $(\Sigma,h)$ and $K\psi_j = \omega_j^2\psi_j$ with $\omega_j>0$. The index set $J$ is countable and we
assume (without loss of generality) that there exists, for each $j\in J$, a (unique) $\bar{j}\in J$
such that $\psi_{\bar{j}}=\overline{\psi_j}$; we allow for the possibility
that $\bar{j}$ might equal $j$. Clearly, $\omega_{\bar{j}}=\omega_j$ for all $j\in J$.

We will consider the class of quasifree pure states of the quantized Klein--Gordon field on $M$ -- equivalently, irreducible Fock space representations of the theory, with the Fock vacuum vector representing the state. Given the simple structure of $M$, the
general theory (see, e.g.,~\cite{Wald_qft}, or~\cite{KayWald:1991}) 
reduces to familiar constructions using mode functions. As usual, the aim is to identify a set of mode solutions to the field equation of the form $T_j(t)\psi_j(\ux)$, which, 
together with their complex conjugates, form a basis for the space of complex valued solutions in a suitable sense. The most general choice is,
of course, $T_j(t) = \alpha_j e^{-i\omega_j t} + \beta_j e^{i\omega_j t}$, for complex-valued $\alpha_j$ and $\beta_j$ (obeying certain
conditions described below). The choice of basis amounts to the
choice of a state: for example, the ultrastatic
vacuum results from the choice $\alpha_j=1$, $\beta_j=0$ for all $j\in J$.
For $\tau<\infty$, the S-J prescription also takes this form~\cite{CF-RV-ultraSJ},\footnote{The S-J prescription
fails in the case $\tau=\infty$. If one takes a suitable limit of S-J states as $\tau\to\infty$, however, the 
ultrastatic ground state is obtained in the limit~\cite{CF-RV-ultraSJ}.}
with
\[
\alpha_j = \sqrt{\frac{\|S_j\|}{\|C_j\|}}\left(1-
\frac{\delta_j}{2}\right),
\qquad \beta_j =\sqrt{\frac{\|S_j\|}{\|C_j\|}}
\frac{\delta_j}{2},
\]
where
$\|S_j\|$
and $\|C_j\|$ denote the $L^2(-\tau,\tau)$ norms of the functions
$S_j(t)= \sin\omega_j t$ and 
$C_j(t)= \cos\omega_j t$, i.e.,
\[
\|C_j\|^2=  \tau (1+\sinc 2\omega_j\tau),\qquad
\|S_j\|^2=  \tau (1-\sinc 2\omega_j\tau),
\]
and
\begin{equation}\label{eq:deltaj}
\delta_j = 1-\frac{\|C_j\|}{\|S_j\|} = 1 - \sqrt{1 + \frac{2\sinc 2\omega_j \tau}{1-\sinc 2\omega_j \tau}}
 = -\sinc 2\omega_j \tau + O((\omega_j\tau)^{-2}).
\end{equation}
Here, we have used the notation $\sinc x = \sin x/x$. 

Returning to the general case, one then constructs a Fock space, in which the field will be given as
\begin{equation}\label{eq:Fock}
\phi(t,\ux) =  \sum_{j\in J} 
\frac{1}{\sqrt{2\omega_j}} 
\left(\left(\alpha_j
e^{-i\omega_j t} +\beta_j
e^{i\omega_j t}\right)\psi_j(\ux) a_j +
\text{h.c.}\right),
\end{equation} 
where $[a_j,a_k^*]=\delta_{jk}\II$ and the Fock vacuum,
annihilated by all $a_j$, will be denoted by $\Omega$.
The corresponding two-point function is given formally by 
\[
W(t,\ux;t',\ux') = \sum_{j\in J} \frac{1}{2\omega_j}
\left(\alpha_j e^{-i\omega_j t} + \beta_j e^{i\omega_j t}\right)
\left(\overline{\alpha_j} e^{i\omega_j t'} + \overline{\beta_j} e^{-i\omega_j t'}\right)\psi_j(\ux)\psi_{\bar{j}}(\ux').
\]

As mentioned above, the $\alpha_j$ and $\beta_j$ are subject
to various conditions. We will require that the series for $W$ 
should converge in the sense of distributions, which means that
if the summands are individually smeared against
$F(t,\ux)G(t',\ux')$ for test functions
$F, G\in \CoinX{(-\tau,\tau)\times\Sigma}$, then the resulting
series should converge in $\CC$, with the sum defining the value
of $W(F,G)$. Furthermore, the commutation relations of the field
require that 
\begin{align*}
W(t,\ux;t',\ux')-W(t',\ux';t,\ux) &= iE(t,\ux;t',\ux') \\ &= \sum_{j\in J} \frac{1}{2\omega_j}
\left(e^{i\omega_j (t-t')}-e^{-i\omega_j (t-t')}\right)\psi_j(\ux)\psi_{\bar{j}}(\ux'),
\end{align*}
which leads to the relations 
\begin{equation}\label{eq:alphas_betas}
|\alpha_j|^2-|\beta_{\bar{j}}|^2 = 1,\qquad \alpha_j\overline{\beta_j} = 
\alpha_{\bar{j}}\overline{\beta_{\bar{j}}} \qquad (j\in J)
\end{equation}
on considering smearings against test functions of the form
$f(t)g(t')\psi_{\bar{j}}(\ux)\psi_j(\ux')$. Taking absolute values in the second set of relations in Eq.~\eqref{eq:alphas_betas}, 
and using the first, we obtain $|\alpha_j|^2(1+|\alpha_{\bar{j}}|^2) = 
|\alpha_{\bar{j}}|^2 (1+|\alpha_j|^2)$ for all $j$ and hence $|\alpha_j| = 
|\alpha_{\bar{j}}|$, $|\beta_j| = 
|\beta_{\bar{j}}|$ for all $j\in J$. As a rephasing 
\[
\alpha_j\mapsto \alpha_j e^{i\chi_j}, \qquad \beta_j\mapsto \beta_j e^{i\chi_j}
\qquad (\chi_j\in\RR)
\]
leaves $W$ invariant, and corresponds to a simple unitary transformation 
on Fock space, we may assume without loss of generality that $\alpha_j\ge 1$
for all $j\in J$, whereupon the class of representations considered
is parameterized by the choice of each $\beta_j\in\CC$ subject to
$\beta_j =\beta_{\bar{j}}$; we then have $\alpha_j=\sqrt{1+|\beta_j|^2}$. As mentioned, it is additionally required that the series for $W$ should converge in the sense of distributions, so the $\beta_j$ cannot be too badly behaved. 
Rather than study the convergence question in detail, we assume
for simplicity that there is a polynomial $Q$ such that $|\beta_j|\le Q(\omega_j)$ for all $j\in J$, which entails a 
similar polynomial bound on the growth of $\alpha_j$. In fact, as we will soon see, the $\beta_j$ must actually decay if the fluctuations of ${:}\dot{\phi}^2{:}(F)$
are to be finite. For simplicity of presentation, we proceed with
formal computation, though our conclusions would also be reached by a more careful analysis. 

Defining normal ordering as usual in the Fock space, we find (formally)
\begin{align*}
{:}\dot{\phi}^2(t,\ux){:} \Omega &= 
-\frac{1}{2}\sum_{j,j'\in J} \sqrt{\omega_j\omega_{j'}} 
\overline{\left(\alpha_j
e^{-i\omega_j t} -\beta_j
e^{i\omega_j t}\right)
\left(\alpha_{j'}
e^{-i\omega_{j'} t} -\beta_{j'}
e^{i\omega_{j'} t}\right)\psi_{j}(\ux)\psi_{j'}(\ux)} 
a_j^* a_{j'}^* \Omega  \\
&= 
-\frac{1}{2}\sum_{j,j'\in J} \sqrt{\omega_j\omega_{j'}} 
\left(\alpha_j
e^{i\omega_j t} -\overline{\beta_j}
e^{-i\omega_j t}\right)
\left(\alpha_{j'}
e^{i\omega_{j'} t} -\overline{\beta_{j'}}
e^{-i\omega_{j'} t}\right) \overline{\psi_{j}(\ux)}\psi_{j'}(\ux)
a_{j}^* a_{\bar{j'}}^* \Omega
\end{align*}
where, in the second step, we have used the fact that the $\alpha_j$ are real, 
and relabelled the $j'$ sum by $j'\mapsto \bar{j'}$. We smear against test functions so that
\[
{:}\dot{\phi}^2{:}(f\otimes g) = \int_{-\tau}^\tau dt\int_\Sigma
\dvol_\Sigma(\ux)\, {:}\dot{\phi}^2(t,\ux){:} f(t)g(\ux),
\]
with $g$ smooth and compactly supported on $\Sigma$, and $f$
smooth and compactly supported in $(-\tau,\tau)$. 

As ${:}\dot{\phi}^2{:}(f\otimes g)$ has vanishing expectation in 
the state $\Omega$, its fluctuation (or dispersion) is simply the 
square root of the expectation of its square; in short, 
the norm of ${:}\dot{\phi}^2{:}(f\otimes g)\Omega$. 
For simplicity, we will take $g(\ux) \equiv 1$ (recall that $\Sigma$ is
compact) and $f$ to be real and even, so that its Fourier transform
\[
\hat{f}(\omega) = \int dt\, f(t) e^{i\omega t}
\]
is also real-valued and even, and decays faster than polynomially
as $\omega\to\infty$. Together with the orthonormality
of the $\psi_j$, our assumptions reduce the sum over $j$ and $j'$ to those terms
for which $j'=j$, giving 
\[
{:}\dot{\phi}^2{:}(f\otimes 1) \Omega
= -\frac{1}{2}\sum_{j\in J} \omega_j 
\left( (\alpha_j^2+\overline{\beta_j}^2)\hat{f}(2\omega_j)
-2
 \alpha_j\overline{\beta_{j}}  \hat{f}(0)
\right)  
a_j^* a_{\bar{j}}^* \Omega ,
\]
where we have also used that $\hat{f}$ is real and even. Hence
\begin{equation}\label{eq:normsq}
\|{:}\dot{\phi}^2{:}(f\otimes 1) \Omega\|^2 
= \frac{1}{2}\sum_{j\in J} \omega_j^2
\left| (\alpha_j^2+\overline{\beta_j}^2)\hat{f}(2\omega_j)
-2
 \alpha_j\overline{\beta_{j}}  \hat{f}(0)
\right|^2.
\end{equation}
While the above computation was rather formal, a more careful
analysis leads to the same identity provided that the right-hand side 
converges; if it diverges, this may be interpreted as showing that
${:}\dot{\phi}^2{:}(f\otimes 1)$ has infinite fluctuations in the state $\Omega$.\footnote{The point is that expressions such as
$\ip{a_j^* a_k^* \Omega}{{:}\dot{\phi}^2{:}(f\otimes 1) \Omega}$
may be defined {\em in the sense of quadratic forms} by a point-splitting
prescription relative to the two-point function $W$; the question
is whether these matrix elements can be interpreted as literal
Fock space inner products, which amounts to the convergence or otherwise of the right-hand side of~\eqref{eq:normsq}.}

In order to estimate this sum, we will use two facts. 
First, general results on the spectrum
of the Laplacian~\cite[Ch.~8, \S 3]{Taylor_volII} permit us to choose
$N\in\NN$ for which 
$\sum_{j\in J} (\omega_j^2+m^2)^{-2N}$ converges. Second, 
as $f$ is smooth and compactly supported, $\sup_{j\in J} |\hat{f}(\omega_j) P(\omega_j)|$ is
finite for any polynomial $P$. Expanding the absolute value in \eqref{eq:normsq},
we find two terms involving $\hat{f}(2\omega_j)$, both of which lead to 
convergent sums on combining the two facts just mentioned with the
polynomial growth bounds on $\alpha_j$ and $\beta_j$. Hence 
${:}\dot{\phi}^2{:}(f\otimes 1) \Omega$
has finite norm if and only if 
\[
\sum_{j\in J} \omega_j^2 |\alpha_j\beta_j|^2 <\infty.
\]
Introducing parameters $\zeta_j=\tanh^{-1}(|\beta_j|/|\alpha_j|)$, 
the condition is equivalent to
\begin{equation}\label{eq:zetas}
\sum_{j\in J} \omega_j^2 \sinh^2 (2\zeta_j)<\infty,
\end{equation}
which implies that $\omega_j\zeta_j\to 0$ as
$j\to\infty$ in some (and hence any) ordering of $J$ by the natural numbers.

In the case of the S-J state, for instance, we have $\alpha_j\sim 1$ and $\beta_j\sim
 -\sinc 2\omega_j \tau$ as $j\to\infty$ (recall that $\tau<\infty$), so
$\omega_j\zeta_j\sim \sin\omega_j \tau$. The following is now essentially immediate,
and provides a further pathology of S-J states that casts substantial doubt
on their physical relevance. 
\begin{Thm} 
For general $(\Sigma,h)$, the set of $\tau\in(0,\infty)$ for which ${:}\dot{\phi}^2{:}$ has finite fluctuations
in the corresponding S-J state is contained in a set of measure zero. If $(\Sigma,h)$ is either a flat
$3$-torus or round $3$-sphere, then there is no $\tau\in (0,\infty)$ for which ${:}\dot{\phi}^2{:}$ has finite fluctuations
in the corresponding S-J state.
\end{Thm}
{\em Proof:} It was shown in~\cite{CF-RV-ultraSJ} that the set of $\tau\in(0,\infty)$ for which $\sin\omega_j \tau\to 0$ 
has vanishing Lebesgue measure for general $(\Sigma,h)$ and is empty in the two special cases mentioned.
${}$ \hfill
$\square$
\\[6pt]
This result can be seen in the same vein as 
the observation made by Brunetti, Fredenhagen and Hollands that the `$\alpha$-vacua' states
on de\,Sitter spacetime yield infinite fluctuations for the Wick-ordered stress-energy tensor,
which casts serious doubts on their utility as physical states
\cite{BruFreHol}. 

Returning to the general case, the above argument is easily modified to deal with Wick squares of higher time derivatives of the field, with consequent increase in the
power of $\omega_j$ in the bound Eq.~\eqref{eq:zetas}. 
Moreover, one sees easily that sufficient decay of the $\zeta_j$ 
implies that ${:}(\partial^k \phi/\partial t^k)^2{:}(f\otimes 1)\Omega$
has finite norm for any $f\in\CoinX{-\tau,\tau}$, not just those that
are real and even. This proves
the following result.
\begin{Thm} \label{thm:equiv1}
In any Fock representation of the field taking the form Eq.~\eqref{eq:Fock},
in which the $\beta_j$ obey a polynomial bound $|\beta_j|\le Q(\omega_j)$
for some polynomial $Q$ and all $j\in J$,  the following are
equivalent:
\begin{enumerate}
\item ${:}(\partial^k \phi/\partial t^k)^2{:}(f\otimes 1)\Omega$ has finite
norm for all $k\in\NN$, where $f\in\CoinX{-\tau,\tau}$;
\item $\sup_{j} |P(\omega_j)\sinh 2\zeta_j|<\infty$ for all polynomials $P$.  
\end{enumerate}
\end{Thm}
In other words, the requirement of finite fluctuations for the above operators
forces $\alpha_j\to 1$ and $\beta_j\to 0$ with rapid convergence (faster than
polynomially in $1/\omega_j$). 

We now show that the state constructed under these conditions
is, in fact, Hadamard. This requires some use of microlocal techniques
to compute the wave-front set $\WF(u)$ of the Hilbert-space-valued
distribution $u(F)=\phi(F)\Omega$.\footnote{The original microlocal
formulation of the Hadamard condition used the two-point function~\cite{Radzikowski_ulocal1996}.} 
The procedure is as follows; references 
include \cite{SVW2002, Sanders:2010}.
Fix any $p=(t_p,\ux_p)$ in the slab and consider local coordinates
$y^i$ ($1\le i\le d-1$) for $\Sigma$ near $\ux_p$ and use $y^0=t$
to give spacetime coordinates $y^\mu(t,\ux)$ ($0\le\mu\le n$) near $p$.\footnote{The
specific choice of coordinates is irrelevant.} Let $k'\in T^*_pM$ be a nonzero covector at $p$, 
with components $k'_\mu$ in the chosen coordinates, and define a function 
$e_{\lambda,k'}(t,\ux) = e^{i\lambda k'_\mu y^\mu(t,\ux)}$ on the $y^\mu$ coordinate chart.
Then the pair $(p,k) \in T^*_pM$ is said 
to be a {\em regular direction} for $u$ if there exists $F\in\CoinX{M}$ with $F(p)\neq 0$ 
such that $\|u(Fe_{\lambda,k'})\|=O(\lambda^{-M})$ as $\lambda\to +\infty$ for every $M\in\NN$
uniformly for all $k'$ in an open neighbourhood of $k$.
The {\em wave-front set} $\WF(u)$ is the set of all $(p,k)\in T^*M$ with $k\neq 0$ that are {\em not} regular directions.
It is known~\cite{SVW2002} that the state $\Omega$ is Hadamard if and only if $\WF(u)\subset \Nc^-$,
the bundle of past-directed null covectors on $M$. 

To start, we will show that every $(p,k)\in T^*M$ with $k_0>0$ is a regular direction for $u$. 
Observe that
\[
\phi(f\otimes g)\Omega = \sum_{j\in J} \frac{1}{\sqrt{2\omega_j}}
\left(\alpha_j \hat{f}(\omega_j) +\overline{\beta_j}\hat{f}(-\omega_j)\right)\ip{\psi_j}{g} a_j^*\Omega
\]
for any $f\in\CoinX{-\tau,\tau}$ and $g\in \CoinX{\Sigma}$, and hence
\begin{equation}\label{eq:phibound}
\|\phi(f\otimes g)\Omega\|^2 \le \|g\|^2 
\sum_{j\in J} \frac{1}{2\omega_j}
\left|\alpha_j \hat{f}(\omega_j) +\overline{\beta_j}\hat{f}(-\omega_j)\right|^2.
\end{equation}
Now fix $f$ and $g$ so that $f(t_p)\neq 0$ and $g(\ux_p)\neq 0$,
with $g$ supported inside the coordinate chart of the $y^i$. 
The inequality~\eqref{eq:phibound} implies 
\begin{equation}\label{eq:est}
\|\phi((f\otimes g)e_{\lambda,k'})\Omega\|^2 \le \|g\|^2 \Upsilon(\lambda k'_0),
\end{equation}
where 
\[
\Upsilon(\mu) = 
\sum_{j\in J} \frac{1}{2\omega_j}
\left|\alpha_j \hat{f}(\mu+\omega_j) +\overline{\beta_j}\hat{f}(\mu-\omega_j)\right|^2
\]
decays faster than any inverse power as $\mu\to +\infty$, assuming that $\beta_j\to 0$ faster than
polynomially in $1/\omega_j$, as will be shown in Appendix~\ref{appx:decay}. Thus for
any $k$ with $k_0>0$, 
the right-hand side of \eqref{eq:est} decays rapidly as $\lambda\to +\infty$, uniformly for
$k'$ in the open neighbourhood $\{\ell\in\RR^4: \ell_0>k_0/2\}$ of $k$; 
it follows that all points of the form $(p,k)\in T^*M$ with $k_0>0$ are regular directions
for the vector-valued distribution $u$, and therefore $\WF(u)\subset \{(p,k)\in T^*M: k_0<0\}$.  
However, as $u$ is a weak solution to the Klein--Gordon equation, its wave-front set must also be contained in the 
characteristic set of the Klein--Gordon operator \cite{DuijHorII}, i.e., the bundle of null co-vectors. Therefore, $\WF(\phi(\cdot)\Omega)\subset \Nc^-$, the bundle of past-directed null covectors, and this establishes that the state $\Omega$ is in fact Hadamard. On the other hand, 
all Wick polynomials have finite fluctuations in Hadamard states by the analysis in~\cite{BrFr2000}
(in any globally hyperbolic spacetime) so we have proved:
\begin{Thm}
The state $\Omega$ is Hadamard if and only if ${:}(\partial^k\phi/\partial t^k)^2{:}$ has finite
fluctuations in $\Omega$ for every $k\in\NN_0$. 
\end{Thm}

\section{Locally pure states and boundary singularities}\label{sect:loc_pure}

The S-J prescription assigns a pure state to bounded regions of spacetime~\cite{AAS,CF-RV-ultraSJ}.
In this section, we explain why any such prescription can be expected to result in singular
behaviour (of a type described below) at the boundary of the region. In fact, the same
applies to all states on the bounded region that are defined by a density matrix in the GNS representation
of a pure state. We work in an algebraic framework of quantum field theory
on $1+d$-dimensional Minkowski space~\cite{Haag} and for the most part, our considerations are
not restricted to the particular example of the free scalar field. 
A brief summary of how the 
Klein--Gordon field fits into the general operator algebraic framework can be found in Appendix~\ref{appx:AQFT}. Our results are consistent
with (but go much further than) the computations performed in~\cite{SorkinGmbH} where the two-point function of the massless Klein--Gordon field is studied near the boundary of a two-dimensional Minkowski diamond. 

The starting assumption is that, to each open relatively compact set $O$ of Minkowski space,
there is a corresponding unital $C^*$-algebra $\Ac(O)$, consisting of the observables of the
theory localised in $O$. These algebras obey the following conditions:
(a) if $O\subset\tilde{O}$ then $\Ac(O)$ is assumed to be a subalgebra
of $\Ac(\tilde{O})$ and they share a common unit; (b) $\Ac(O)$ and $\Ac(\tilde{O})$ are assumed to commute elementwise if $O$ and $\tilde{O}$ are spacelike separated. We will not make any assumptions about the action of spacetime symmetries at this level.  

Let $M$ be a double-cone in Minkowski space, with corresponding algebra $\Ac(M)$. Recall that
a state on $\Ac(M)$ is a linear map $\omega:\Ac(M)\to\CC$ that is normalised ($\omega({\bf 1})=1$)
and positive ($\omega(A^*A)\ge 0$ for all $A\in\Ac(M)$). Any state on $\Ac(M)$
induces a Hilbert space GNS representation; this representation is irreducible if and only if 
the state is pure~\cite[II.6.4.8]{Blackadar}. Further, any state $\omega$ on $\Ac(M)$
induces a whole class of states that are defined by density matrices in the GNS representation
of $\omega$; these states are said to be {\em normal} to $\omega$ (in particular, this includes all vector
states in the representation, and $\omega$ itself). 

We wish to show that no state normal to a pure state of the algebra $\Ac(M)$ can be extended to
any strictly larger double cone $N$ without encountering singular behaviour at the boundary
of $M$. Our argument actually proceeds in reverse, by
showing that singularity-free states of $\Ac(N)$ do not restrict to $\Ac(M)$
as pure states, or even states that are normal to pure states. 

Let $\omega$ be any state of $\Ac(N)$. It induces a GNS representation
$(\Hc,\pi,\Omega)$ of $\Ac(N)$ so that $\ip{\Omega}{\pi(A)\Omega} = 
\omega(A)$ for all $A\in\Ac(N)$, with $\Omega$ as a cyclic vector for
the representation. 
Thus, a state $\omega'$ of $\Ac(N)$ is normal to $\omega$ if there is a density matrix
$\varrho$ on $\Hc$ such that $\omega'(A) = {\rm Tr}(\varrho \pi(A))$
holds for all $A \in \Ac(N)$.
We assume, as a requirement on $\omega$, 
that the Hilbert space $\Hc$ is separable. If $O$ is any open bounded region contained
in $N$, we obtain a subalgebra $\pi(\Ac(O))$ of the bounded 
operators $\Bc(\Hc)$ of $\Hc$, and a corresponding von Neumann algebra $\Nc(O):=\pi(\Ac(O))''$,
where the prime denotes the operation of forming the commutant.\footnote{The commutant
of any subset $\mathcal{X}$ of $\Bc(\Hc)$ is defined as
$\mathcal{X}' = \{ B \in \Bc(\Hc) : AB = BA\ \ \text{for all} \ A \in 
\mathcal{X}\}$, and the bicommutant is defined as $\mathcal{X}'' = (\mathcal{X}')'$.} 
By von Neumann's bicommutant theorem, $\Nc(O)$ coincides with the weak closure of $\pi(\Ac(O))$ \cite[I.9.1.1]{Blackadar}.

To explain the singular behaviour we have in mind, we assume
in addition that the theory has a description involving field operators. 
Specifically, we assume that there is a dense domain $\Dc\subset\Hc$
containing the GNS vector $\Omega$ 
and, to every real-valued test function $f\in\CoinX{N}$, there is a corresponding hermitian
smeared field operator $\phi(f)$ leaving $\Dc$ invariant, depending
linearly on $f$ and so that the operator closures $\overline{\phi(f)}$
exist and are affiliated to the local algebras.\footnote{That is, when one takes a polar decomposition
$\overline{\phi(f)}=U_f|\overline{\phi(f)}|$, the partial isometry
$U_f$, and every bounded function of $|\overline{\phi(f)}|$, belong to $\Nc(O)$ for any $O$ containing the support of $f$.}  
We may form $n$-point functions of the field $\phi$ in state $\omega$ by 
setting
\[
W_\omega^{(n)}(f_1,\ldots,f_n)= \ip{\Omega}{\phi(f_1)\cdots
\phi(f_n)\Omega}
\]
for any test functions $f_k\in \CoinX{N}$ and we require these
to determine distributions on $N$. 

The next step is to consider scalings of the $n$-point functions. 
Let $p$ be a point in $N$ and define the scaling maps
$$ \beta_{p,\lambda} f(x) = f((x - p)/\lambda)\,, \quad f \in C_0^\infty(\mathbb{R}^{1+d})\,,\ x \in N\,,\ 1 > \lambda > 0 $$
which contract the supports of test-functions to the point $p$ as $\lambda \to 0$. The following definition is based on a condition
in~\cite[\S 16.2.4]{BaumWolle:1992}.
\begin{Def} The state $\omega$ has a {\em regular scaling limit} at $p$ if, in addition to the conditions above, the following two
properties hold:\\[2pt]
${}$ \quad (i) The intersection of the local von Neumann algebras in the GNS representation induced by $\omega$ is trivial at $p$,
i.e.\
$$ \bigcap_{O \owns p} \mathcal{N}(O) = \mathbb{C}{\bf 1} \,;$$
${}$ \quad (ii) There is some monotone, positive-valued function $\lambda \mapsto \nu(\lambda)$ so that the limits
$$ W_{0}^{(n)}(f_1,\ldots,f_n) = \lim_{\lambda \to 0} \, \nu(\lambda)^n W^{(n)}_\omega(\beta_{p,\lambda}f_1,\ldots,\beta_{p,\lambda}f_n)
  $$
exist for all $n \in \mathbb{N}$ and all $f_j \in C_0^\infty(\mathbb{R}^{1+d})$ and define (non-trivial) $n$-point functions satisfying the Wightman axioms for a
quantum field theory\footnote{This `scaling limit theory' will typically differ from the original theory~\cite{Buch:1996}.} in its vacuum representation~\cite{StreaterWightman}.
\end{Def}
The existence of regular scaling limits is expected for quantum field theories possessing an ultraviolet fixed point~\cite{Buch:1996}. We regard the failure of a regular scaling limit as indicating a pathology of the state $\omega$. For our purposes, the importance of scaling limits is that
they permit us to determine the {\em type} of the local von Neumann algebras. We will only need to consider algebras of types I and III. 
A von Neumann algebra $\Nc$ is said to be of type I if it is isomorphic
[as a von Neumann algebra] to the algebra of bounded operators
on some Hilbert space. A von Neumann algebra $\Nc$ is of type III 
if it contains no finite projections, which can also be expressed as follows: If $P \in \Nc$ is any 
non-zero projection
operator, then there is a proper sub-projection $Q \in \Nc$ (i.e.\ $QP = PQ = Q$ and $Q \ne P$) together
with an operator $V \in \Nc$ such that $V^*V = P$ and $VV^* = Q$.  
 The type is preserved under von Neumann algebra isomorphisms. Moreover, 
in cases where the von Neumann algebra is obtained by
taking the weak closure of a representation of a $C^*$-algebra,
we describe the representation as having the same type as the
von Neumann algebra. 

The following is a restatement of~\cite[Theorem~16.2.18]{BaumWolle:1992}, 
which is a development of a seminal result of Fredenhagen~\cite{Fredenhagen:1985}. 
\begin{Prop} \label{prop:SLtype3}
Let $M$ be a double cone with compact closure contained in $N$ and let $p$ be a point in the spacelike boundary of $M$. 
Suppose that the state $\omega$ of $\Ac(N)$ has a regular scaling limit at $p$. Then
the local von Neumann algebra $\mathcal{N}(M)=\pi(\Ac(M))''$ is of type III, i.e., 
the restriction $\pi|_{\Ac(M)}$ is a type III representation of $\Ac(M)$. 
\end{Prop} 
(In fact, even type III${}_1$ is proved in the references given.)
Our result is a simple consequence.
\begin{Cor} \label{Cor:type3notpure}
(a) Under the above conditions, the GNS representation
induced by $\omega|_{\Ac(M)}$ is of type III. In particular, 
$\omega|_{\Ac(M)}$ is neither a pure state, nor is it normal to the GNS representation of a pure state of $\Ac(M)$. (b) If, in addition, the von Neumann algebra 
$\mathcal{N}(M)$ is a factor\footnote{That is, $\mathcal{N}(M)$
intersects its commutant only in multiples of the identity operator.}
and $\omega'$ is normal to $\omega$, then $\omega'|_{\Ac(M)}$ 
also induces a type III representation. 
\end{Cor}
{\noindent\em Proof.} (a) The restriction $\pi|_{\Ac(M)}$ defines a representation of $\Ac(M)$ on $\HH$
such that $\omega|_{\Ac(M)}(A)=\ip{\Omega}{ \pi|_{\Ac(M)}(A)\Omega}$ for all $A\in\Ac(M)$. 
Defining $\HH'$ to be the closure of $\pi|_{\Ac(M)}(\Ac(M))\Omega$, it follows that
$(\HH',\pi|_{\Ac(M)}|_{\HH'},\Omega)$ has all the properties of the GNS representation
of $\Ac(M)$ induced by $\omega|_{\Ac(M)}$ and may therefore be identified with it.
Thus, the GNS representation induced by $\omega|_{\Ac(M)}$ may be identified with a subrepresentation of $\pi|_{\Ac(M)}$. 
Applying Prop.~\ref{prop:SLtype3} and \cite[III.5.1.7]{Blackadar}, we have shown that $\omega|_{\Ac(M)}$ induces a type III representation. 

On the other hand, any pure state $\omega_0$ of $\Ac(M)$ induces an irreducible representation, which is
therefore of type I. Irreducibility implies that $\omega_0$ is {\em primary};\footnote{A state of a $C^*$-algebra is primary if the
von Neumann algebra formed in the corresponding GNS representation
is a factor.} by \cite[10.3.14]{KadRing_ii}, all
its normal states induce GNS representations 
that are quasi-equivalent to that of $\omega_0$, so they therefore 
have the same type \cite[III.5.1.7]{Blackadar}. Hence 
any (state normal to a) pure state of $\Ac(M)$ has a type I representation. 

(b) As $\mathcal{N}(M)$ is a factor, it follows, first, that $\omega'|_{\Ac(M)}$ is normal to $\omega|_{\Ac(M)}$ (see, e.g., Appendix b) of \cite{BrFrVe03}) and, second, that $\omega|_{\Ac(M)}$ is primary (this is a consequence e.g.\ of \cite[I.9.1.5]{Blackadar}). By the argument
just used in part (a), $\omega'|_{\Ac(M)}$ induces a representation
of the same type as $\omega|_{\Ac(M)}$, i.e., type III. 
${}$ \hfill $\square$
 \\[6pt]
Turning this around, if a (state normal to a) pure state of $\Ac(M)$ is extended to the algebra of 
any larger region $N$ then the extended state cannot be associated with a regular scaling limit at
{\em any} point of the spacelike boundary of $M$. 
Note that although we have described scaling limits in terms of fields, there is an intrinsically operator-algebraic version of these results which
would lead to essentially the same conclusion~\cite{BuchVer:1995}.

Let us now apply these results for the case of the linear scalar field with particular regard to S-J states for
double cones. Therefore, $\Ac(O)$ will now denote the $C^*$-Weyl algebra of the linear scalar field 
defined for a double cone $O$ in Minkowski spacetime
(of arbitrary dimension, not less than 2); the formal definition can be found in Appendix \ref{appx:AQFT}.
By $\omega_{SJ,O}$ we denote
the S-J state for the double cone $O$, i.e.\ the quasifree state on $\Ac(O)$ determined by the
two-point function $W_{SJ}$ of \eqref{eq:DefSJ}, for the case that the spacetime with respect to which $W_{SJ}$
is defined is the double cone $O$ with the induced Minkowski metric. 
\begin{Prop} \label{Prop:SJnastyonbdry} ${}$ \\[2pt]
(a) \ \ \ Let $M \subset N$ be a strict inclusion of double cones (so that $\overline{M} \subset N$, where both $M$ and $N$
are open double cones), and let $\omega_{SJ,M}$ be the S-J state for the smaller double cone $M$. Then neither $\omega_{SJ,M}$, nor
any state normal to it, can admit an  
extension to a Hadamard state, or in fact any state normal to a Hadamard state, on $\Ac(N)$.
\\[6pt]
(b) \ \ \ Supposing again that $M \subset N$ is a strict inclusion of double cones, then the following statements are mutually
exclusive.
\begin{itemize}
 \item[(i)] $\omega_{SJ,N}$ has a regular scaling limit
 at some point $p$ in the spacelike boundary of $M$.
 \item[(ii)] $\omega_{SJ,M}$ (or a state normal to it) extends to a state $\omega$ on $\Ac(N)$ which is normal to $\omega_{SJ,N}$.
\end{itemize}

\end{Prop}
{\noindent\em Proof.} ${}$ \\[2pt]
(a) Suppose first that $\omega$ is a Hadamard state on $\Ac(N)$.
Since Hadamard states possess regular scaling limits at all points in spacetime \cite{SahlVer-rmp}, Corollary
\ref{Cor:type3notpure}(a) implies that $\omega|_{\Ac(M)}$
induces a type III representation. Morever, the von Neumann algebra $\Nc(M)$ [formed, as above, with respect to the GNS-representation of $\omega$], is a factor \cite{Verch:1996rmp}. Applying  Corollary
\ref{Cor:type3notpure}(b), if $\omega'$
is any state normal to $\omega$, then $\omega'|_{\Ac(M)}$ induces
a type III representation and cannot coincide with the S-J state $\omega_{SJ,M}$, which is pure, or any state normal to it.
\\[6pt]
(b) If $\omega_{SJ,N}$ has a regular scaling limit at any point in the spacelike boundary of $M$,
then the von Neumann algebra $\Nc_{SJ,N}(M) = \pi_{SJ,N}(\Ac(M))''$ formed in the GNS representation of
$\omega_{SJ,N}$ is of type III. On the other hand, since $\omega_{SJ,N}$ is a pure state, and thus primary, any state $\omega$ on $\Ac(N)$ which is normal to $\omega_{SJ,N}$ is
quasiequivalent to it by~\cite[10.3.14]{KadRing_ii}. This implies that there is an isomorphism
$\phi: \pi_{SJ,N}(\Ac(N))'' \to \pi_\omega(\Ac(N))''$ between the von Neumann algebras formed in the
GNS representations of the respective states which also satisfies $\phi \circ \pi_{SJ,N} = \pi_\omega$.
Consequently, $\phi$ restricts to a von Neumann algebra isomorphism between $\pi_{SJ,N}(\Ac(M))''$ and
$\pi_\omega(\Ac(M))''$, and therefore the latter von Neumann algebra is of type III, too. By arguments used in Corollary~\ref{Cor:type3notpure}(a), it follows that $\omega|_{\Ac(M)}$ has a GNS representation of 
type III, and cannot coincide with $\omega_{SJ,M}$ (or any state normal to it) since the latter state is pure.
${}$ \hfill $\Box$
\\[10pt]
{\it Remarks.}
\\[2pt]
(A) The statements of Prop.\ \ref{Prop:SJnastyonbdry} carry over to curved spacetime, upon making the following modifications:
$N$ is a globally hyperbolic spacetime such that the S-J state is defined for the quantized linear scalar field on $N$
(e.g.\ if $N$ is isometrically embedded into a larger spacetime, see \cite{CF-RV-ultraSJ}), and $M$ is globally hyperbolic 
sub-spacetime of $N$ having a sufficiently regular spacelike boundary,
and whose closure is properly contained in $N$. A class of sub-spacetimes $M$ having the
required properties is e.g.\ given by those of the form $M = {\rm int}(D(B))$ where $D(B)$ denotes the domain of
dependence of $B$, and $B$ is a coordinate ball of any Cauchy surface $\Sigma$ of $N$ (with $\Sigma \backslash B$ having 
a non-void open interior). This is based on the following
facts: (i) Proposition \ref{prop:SLtype3} generalizes to curved spacetimes by \cite{Wollenb},\cite[Theorem~16.2.18]{BaumWolle:1992}
(also in purely operator-algebraic setting, see \cite{Verch:SL-CST}),
(ii) Hadamard states
of the quantized linear scalar field have regular scaling limits also in curved spacetime \cite{SahlVer-rmp} and (iii) the
local von Neumann algebras in their GNS representations are factors for spacetime regions $M$ of the said
form \cite{Verch:1996rmp}. 
\\[6pt]
(B) The results clearly indicate that a state which is pure on the algebra of observables of a double cone has
a singular behaviour at the double cone's spacelike boundary in the sense that the state cannot be extended
to an ambient spacetime having regular scaling limits at the spacelike boundary of the double cone. Statement
(b) points at problems if one were to take the point of view that physical states are those which are normal
to S-J states (the natural definition of physical states once one
is given candidate ``vacuum states'' \cite{Haag}). Then the set of physical states depends sensitively on the 
particular choice of a double cone
--- similar to the findings we made for S-J states for ultrastatic slab spacetimes in \cite{CF-RV-ultraSJ} --- or
there are points in spacetime where S-J states fail to have regular scaling limits. However, as mentioned before,
that would commonly be viewed as a pathological short-distance behaviour of a state, particularly for
a linear quantized field.
\\[6pt]
(C)
The type III property of the local von Neumann algebras of observables is a typical feature of relativistic quantum field
theory which is not encountered in quantum mechanics.  A consequence (in combination with the assumption of regular scaling limits)
is that the von Neumann algebras of local observables associated with tangent double cones don't admit product
states which are (locally) normal to the vacuum \cite{SummersWernerAHP49}. It also makes the concept of local particle
number operators problematic, see e.g.\ \cite{Summers:SubIndep} and literature cited there for discussion.

\section{Further remarks on S-J States}\label{sect:further_SJ}

In this section we make some further observations about the
specific case of S-J states. First, we consider whether the failure of
S-J states on ultrastatic slab spacetimes, which we have discussed in this paper and in \cite{CF-RV-ultraSJ}, 
can be removed by `softening the boundary' of the slab by an averaging
procedure. Second, we show that the S-J states on such spacetimes (and their averaged
variants) fail to be locally Hadamard. We mention, however, that recently Brum and Fredenhagen
have considered a somewhat different method of ``softening the boundary'' of the ultrastatic slab,
leading to an altered definition of S-J states which turns out to result in Hadamard states \cite{BrumFredenhagen:2013}.
While this result indicates that it is really the ``sharp boundaries'' of an ultrastatic slab in time-direction
which causes the failing Hadamard property of the associated S-J states, our present results show that Hadamard
states cannot be gained by straightforward averaging procedures on the S-J states of ultrastatic slabs.

\subsection{Averaging procedures}

One could imagine many different prescriptions: here, we
study (a) the effect of averaging with respect to a shift
in the slab's absolute time and (b) the effect of averaging with respect to the slab duration parameter $\tau$. As we will see, neither prescription results in a Hadamard state.

We recall from \cite{CF-RV-ultraSJ} that, on a slab spacetime $M=(-\tau,\tau)\times \Sigma$ as in Section~\ref{sect:fluctuations}, the normal ordered SJ two-point function is
\begin{equation}\label{eq:WS_normord}
{:}W_{SJ}{:}(t,\ux;t',\ux') = \sum_{j\in J} \left\{\frac{\delta_j^2\cos\omega_j (t-t')}{4\omega_j(1-\delta_j)}
+\frac{\delta_j(2-\delta_j)}{4\omega_j(1-\delta_j)}\cos\omega_j(t+t')\right\}
\psi_j(\ux)\overline{\psi_j(\ux')},
\end{equation}
where normal ordering is performed with respect to the standard ultrastatic ground state (which is Hadamard) and the $\delta_j$ are given in~\eqref{eq:deltaj}. In \cite{CF-RV-ultraSJ}, we showed
that ${:}W_{SJ}{:}$ cannot be smooth except, perhaps, for a
set of $\tau$ with measure zero. 

For our first averaging prescription, let $\rho\in\CoinX{-\epsilon,\epsilon}$ be a nonnegative and even function, with $\epsilon\ll \tau_0$ and  $\int d\tau \rho(\tau)=1$. We consider the quasifree state with two-point function
\[
\widehat{W}_{SJ}(t,\ux;t',\ux') = \int d\tau\rho(\tau) W_{SJ,\tau_0}
(t+\tau,\ux;t'+\tau,\ux')
\]
on the slab
$M=(-\tau_0+\epsilon,\tau_0-\epsilon)\times\Sigma$. 
As the ultrastatic ground state is time-translation invariant, a
simple computation gives
\[
{:}\widehat{W}_{SJ}{:}(t,\ux;t',\ux') = \sum_{j\in J} \left\{\frac{\delta_j^2\cos\omega_j (t-t')}{4\omega_j(1-\delta_j)}
+\frac{\delta_j(2-\delta_j)}{4\omega_j(1-\delta_j)}\hat{\rho}(2\omega_j)\cos\omega_j(t+t')\right\}
\psi_j(\ux)\overline{\psi_j(\ux')}.
\]
In order to show that the new state is not globally Hadamard, it
is enough to show that this normal ordered two-point function is not smooth. The argument is much as in \cite[\S 4.2]{CF-RV-ultraSJ}. If ${:}\widehat{W}_{SJ}{:}$ were smooth, then the derivative $\partial^{4k}{:}\widehat{W}_{SJ}{:}/\partial t^{2k}\partial t^{\prime 2k}$ 
would have to be the integral kernel of a Hilbert--Schmidt operator on $L^2((-\tau_0-2\epsilon,
\tau_0+2\epsilon)\times\Sigma)$ for all $k\in\NN$. Given that
$\hat{\rho}(2\omega_j)$ decays rapidly as $j\to\infty$, it is
easily seen that our condition requires that
\[
\sum_{j\in J}  \frac{\omega_j^{4k}\delta_j^2\cos\omega_j (t-t')}{4\omega_j(1-\delta_j)}
\psi_j(\ux)\overline{\psi_j(\ux')}
\]
is also the integral kernel of such a Hilbert--Schmidt operator. 
Proceeding as in \cite[\S 4.2]{CF-RV-ultraSJ} this implies
\[
\sum_{j\in J}  \frac{\omega_j^{8k-2}\delta_j^4 }{(1-\delta_j)^2} <\infty
\]
Given that $\delta_j\to 1$ we deduce, taking $k=1$, that 
$\omega_j\delta_j \to 0$ as $j\to\infty$. This can happen
for at most a measure zero set of $\tau_0$  (see the proof of 
~\cite[Theorem 4.2]{CF-RV-ultraSJ}). 

Turning to the second averaging prescription, let $\rho\in\CoinX{\tau_0-\epsilon,\tau_0+\epsilon}$ be
nonnegative, with $\epsilon\ll \tau_0$ and $\int d\tau \rho(\tau)=1$. We consider the quasifree state with two-point function
\[
\widetilde{W}_{SJ}(t,\ux;t',\ux') = \int d\tau\rho(\tau) W_{SJ,\tau}
(t,\ux;t',\ux')
\]
which is easily seen to be the two-point function of a state on a slab
$M=(-\tau_0+\epsilon,\tau_0-\epsilon)\times\Sigma$. Here,
$W_{SJ,\tau}$ is the two-point function of the SJ state on the slab $(-\tau,\tau)\times\Sigma$ (cf.~\cite[\S 4.5]{CF-RV-ultraSJ}). 
The normal-ordered two-point function is
\[
{:}\widetilde{W}_{SJ}{:}(t,\ux;t',\ux') = \sum_{j\in J}  \left\{
A_j \cos\omega_j (t-t') + B_j \cos\omega_j(t+t')\right\}
\psi_j(\ux)\overline{\psi_j(\ux')}, 
\]
where
\[
A_j = \int d\tau \rho(\tau) \frac{\delta_j(\tau)^2}{4\omega_j(1-\delta_j(\tau))}, \qquad
B_j = \int d\tau \rho(\tau)\frac{\delta_j(\tau)(2-\delta_j(\tau))}{4\omega_j(1-\delta_j(\tau))}
\]
and we have written the $\tau$-dependence of $\delta_j$ explicitly.
It is elementary, using the Riemann--Lebesgue lemma, to show that
\[
A_j \sim \frac{\text{const}}{\omega_j^3}\sim  -B_j 
\]
Following a similar integral kernel argument to the one just sketched, we see
that ${:}\widetilde{W}_{SJ}{:}$ cannot be smooth
(for any $\tau_0$ or $\rho$); accordingly, $\widetilde{W}_{SJ}$ is not globally Hadamard.

\subsection{Failure to be locally Hadamard}

Our results in \cite{CF-RV-ultraSJ} and in the previous section showing that S-J states
and some ``smoothed'' variants cannot be globally Hadamard 
on ultrastatic slabs can be strengthened to concluding that they are not even
locally Hadamard by applying Radzikowski's local-to-global theorem 
for smooth differences of 2-point functions for the KG-field \cite{Radzikowski:loc-to-glob},
which can be re-stated as follows (where also the propagation-of-singularities 
theorem \cite[Sec.\ 6.1]{DuijHorII} is implicitly used):
\begin{Prop} \label{prop:locH}
Let $W$ be the two-point function of a state defined for a globally
hyperbolic spacetime $M$ (assuming $W$ is a bi-distribution). 
Suppose that $W$ is {\em locally Hadamard} in the following sense:
there is a Cauchy-surface $\Sigma$ for $M$, an open neighbourhood
$N$ of $\Sigma$ and an open covering $\{O_k\}_{k \in \mathcal{K}}$ of $N$ (where
$\mathcal{K}$ is a suitable index set) such that $W | C_0^\infty(O_k \times O_k)$
is of Hadamard form for all $k \in \mathcal{K}$.
 Then $W$ is globally Hadamard. 
\end{Prop}

This statement has two Corollaries which show that generically the S-J two-point functions
$W_{SJ,\tau}$
of the KG-field
on ultrastatic slab spacetimes $(-\tau,\tau) \times \Sigma$ cannot be locally Hadamard. The first is an immediate consequence.

\begin{Cor}\label{cor:locH1}
Let $\tau >0$ be such that $W_{SJ,\tau}$ is not globally Hadamard, let $\tilde{\Sigma}$
be any Cauchy-surface for the ultrastatic slab spacetime $(-\tau,\tau) \times \Sigma$,
and let $\{O_k\}_{k \in \mathcal{K}}$ be an open cover of (an open neighbourhood of)
$\tilde{\Sigma}$. Then there must be some indices $\bar{k}$ in $\mathcal{K}$
such that $W_{SJ,\tau} | C_0^\infty(O_{\bar{k}} \times O_{\bar{k}})$ fails to be Hadamard.
\end{Cor}

The statement becomes considerably sharper if the model Cauchy surface $\Sigma$
of the ultrastatic slab spacetime $(-\tau,\tau) \times \Sigma$ carries a transitive group
$G$ of isometries.

\begin{Cor}\label{cor:locH2}
Let $\tau >0$ be such that $W_{SJ,\tau}$ is not globally Hadamard, and assume that $\Sigma$
possesses a transitive group $G$ of isometries. Then for any open subset $O$ of 
$(-\tau,\tau) \times \Sigma$, 
$W_{SJ,\tau} | C_0^\infty(O\times O)$ fails to be Hadamard.
\end{Cor}
{\noindent\em Proof:} To see this, note that we may without restriction of generality assume that 
$O = {\rm int}(D(\{\tau_0\} \times S))$ for some open subset $S$ of $\Sigma$ and
some $\tau_0 \in  (-\tau,\tau)$, as open sets of this type form a base for
the topology of an ultrastatic slab. Then we can move $S$ and hence $O$ around with 
the group $G$ 
and produce an open cover of $\{\tau_0\} \times \Sigma$ in this way. On the other hand,
by construction, $W_{SJ,\tau}$ is invariant under this group action. Thus, assuming that
$W_{SJ,\tau} | C_0^\infty (O \times O)$ were Hadamard, we would conclude that
$W_{SJ,\tau}$ is locally Hadamard on an open cover 
of a Cauchy-surface in the sense of Proposition~\ref{prop:locH} and thus
globally Hadamard, which again results in a contradiction. ${}$ \hfill $\square$
\\[6pt]
The final observation is entirely obvious.
\begin{Cor}
The statements of Cor.~\ref{cor:locH2} and Cor.~\ref{cor:locH2} hold likewise for the smoothed-by-averaging
versions of the S-J two-point functions on ultrastatic slabs.
\end{Cor}

\section{Discussion}

The Hadamard property for states of quantized linear fields on curved spacetimes is well-motived
by various considerations, e.g.\ its intimate connection to quantum energy inequalities and
stable thermodynamical behaviour. Moreover, it permits a systematic and, more
importantly, local and covariant definition of renormalized quantities such as the stress-energy tensor
and Wick-ordered and time-ordered product. These properties are instrumental for setting up a 
local covariant perturbative construction of interacting quantum fields which has hence been
carried out, see e.g.\ \cite{BrFr2000,Ho&Wa01,Ho&Wa02}. In the present paper, we have shown that, for ultrastatic slabs, the Hadamard property
is itself a consequence of imposing finite variance for Wick squares of time-derivatives of
field operators --- a prerequisite for the quantum mechanical interpretation. 
Therefore, the failure of S-J states to be Hadamard states cannot, in our view, be taken lightly.
In particular, the divergence of the variance of the Wick square of the field's time-derivative for
S-J states on ultrastatic slab spacetimes is a strong argument against considering such states as 
physical, or even as surrogate vacuum states. We have also seen, as a consequence of more
general, model-independent arguments that S-J states defined for double cones must be singular
at the spacelike boundary in the sense of not admitting extensions to states on larger spacetime
regions having regular scaling limits at the spacelike boundary of the double cones. Furthermore,
in the ultrastatic slab situation, the Hadamard property cannot be reached at by simple averaging
procedures on S-J states. The altered definition suggested in \cite{BrumFredenhagen:2013} is different
and indicates, in conjunction with our results for double cones, that the problems of S-J states
defined for extendible spacetimes stem prominently from a `sharp cut-off' at the `edges' of a
spacetime. 

The following can, in our view, be concluded from the results. 
Firstly, the Hadamard property is essentially inevitable for physical states of linear quantized fields on a curved spacetime. Secondly, the
definition of S-J states for extendible spacetimes suffers from ultraviolet problems
at the `spacetime edges'. There is an interesting aspect to it, however. It might be that
S-J states (supplemented by a limiting procedure) yield Hadamard states on inextendible, geodesically complete spacetimes, while
the Hadamard property fails for S-J states generically on geodesically incomplete spacetimes.
If a variant of such a statement could be established, then the deviation of S-J states from
the Hadamard property can be taken as a signal for geodesic incompleteness, and at the level
of quasi-free representations of the Weyl-algebra, one can associate an algebraic index-like quantity to it.
This would be of interest also because failure of geodesic completeness is an indication for
singularities in general relativity, and the possibility of associating such a situation with a kind of index
of a physically interpretable quantum field theory on the spacetime in question has attractive features from
a mathematical point of view, potentially with wider implications.

Finally, the definition of S-J states is a continuum extrapolation of a proposed
vacuum-like state for a system of discrete degrees of freedom within the
`causal sets' programme \cite{Johnston:2009,Sorkin:2011}, a certain approach to quantum gravity. There
may be several reasons why this extrapolation does not always lead to Hadamard states.
One origin may be that the S-J prescription for continuum quantum fields is simply not an appropriate continuum limit of the discrete theory. For instance, in the light of our second conclusion, it might be that the limiting procedure ought to exclude geodesically incomplete spacetimes. Another possibility worth contemplating is that the causal set framework might not yet possess sufficient structure to uncover the features relevant for physical
states in a continuum limit, like the microlocal spectrum condition
(or any other expression of dynamical stability roughly asserting that physical states are mixtures of `small perturbations of a ground state or thermal equilibrium state'). Therefore, it might be that our negative
results concerning S-J states indicate that there is some structural element missing
in the causal sets programme which pertains to information about the dynamics.

\vspace{0.5cm}

{\noindent\em The work reported here was partly conducted during the workshop 
``New Trends in Algebraic Quantum Field Theory'', Frascati, September 2012, and the authors thank 
the organisers for financial support. We are also grateful to Klaus Fredenhagen for 
suggesting the idea of computing the fluctuations of Wick polynomials in S-J states.
RV thanks the Department of Mathematics of the University of York for hospitality and financial support on occasion of a visit in
May 2013 during which major work on the manuscript was done. We also thank Felix Finster, Jorma Louko and Rafael Sorkin
for comments.}

\appendix

\section{The Quantized Linear Scalar Field in the Operator Algebraic Setting}
\label{appx:AQFT}

Let us now explain how the linear Klein-Gordon field fits into the framework described in 
Section~\ref{sect:loc_pure}. Let $(M,g)$
be a globally hyperbolic spacetime, for example (a globally hyperbolic subspacetime of) Minkowski space. Then there are uniquely defined  
advanced ($-$) and retarded ($+$) Green's operators $\Ett^\pm$, defined on
$C_0^\infty(M,\mathbb{R})$, for the Klein-Gordon operator
$\Box + m^2$, where $\Box$ is the d'Alembert operator on $(M,g)$ and $m \ge 0$ is a 
constant, and $\supp \Ett^\pm f\subset J^\pm(\supp f)$. (See e.g.,~\cite{BarGinouxPfaffle} for details,
though our conventions differ.) Their difference $\Ett =
\Ett^- - \Ett^+$ is the causal Green's operator. Factorizing the space $C_0^\infty(M,\mathbb{R})$
by the kernel of $\Ett$ gives the real vector space $K = C_0^\infty(M,\mathbb{R})/{\rm ker}(\Ett)$
of equivalence classes $[f] = f + {\rm ker}(\Ett)$ ($f \in C_0^\infty(M,\mathbb{R})$) and
it is known that 
\[
 \sigma([f],[h]) = \int_M f(p) (\Ett h)(p)\,d{\rm vol}_M(p)
\]
supplies a symplectic form on $K$ (where $d{\rm vol}_M$ is the metric-induced volume
form of $(M,g)$). Therefore, $(K,\sigma)$ is a symplectic space. To any symplectic space
there is uniquely associated its {\it Weyl algebra} $\mathcal{W}(K,\sigma)$, the unique
$C^*$ algebra generated by a unit {\bf 1} and a family of elements ${\sf W}([f])$, $[f] \in K$,
subject to the relations
\[
 {\sf W}([0]) = {\bf 1}\,, \quad {\sf W}([f])^* = {\sf W}(-[f])\,, \quad
  {\sf W}([f]){\sf W}([h]) = {\rm e}^{i\sigma([f],[h])/2}{\sf W}([f] + [h])\,,
\]
which are the canonical commutation relations in exponentiated form, also called the Weyl relations.
One can now set $\Ac(M) = \mathcal{W}(K,\sigma)$ and define $\Ac(O)$ as the
$C^*$-subalgebra generated by all ${\sf W}([f])$ with ${\rm supp}(f) \subset O$. The
 assignment $(M,g) \to \Ac(M)$ is, in fact, functorial~\cite{BrFrVe03}.

A two-point function is a distribution $W$ in $\mathscr{D}'(M \times M)$ with the
property that $W(\bar{f},f) \ge 0$\footnote{We write $W(f,h)$ instead of the technically more correct
$W(f \otimes h)$.} for all test-functions $f \in C_0^\infty(M)$ and such that there is
some real scalar product $\mu$ on $K$ with
\[
 W(f,h) = \mu([f],[h]) + \mbox{$\frac{i}{2}$}\sigma([f],[h])
\]
for all real-valued test-functions $f,h \in C_0^\infty(M,\mathbb{R})$, and
extension by requiring complex linearity in both entries of $W$.
Any two-point function $W$ determines a quasifree state $\omega$ on $\Ac(M) = \mathcal{W}(K,\sigma)$
by setting
\[
 \omega({\sf W}([f])) = {\rm e}^{-W(f,f)/2} \quad (f \in C_0^\infty(M,\mathbb{R}))
\]
and extension by linearity. For such a quasifree state $\omega$, one obtains quantum field
operators $\phi(f)$ in the corresponding GNS representation $(\Hc,\pi,\Omega)$ as the 
generators of the $1$-parameter unitary groups $t\mapsto \pi({\sf W}(t [f]))$. 
That is, to each $f\in  C_0^\infty(M,\mathbb{R})$ there is a self-adjoint operator
$\phi(f)$ such that
\[
 \phi(f) \psi= \left. \frac{1}{i}\frac{d}{dt} \pi({\sf W}(t [f]))\psi\right|_{t=0}
\]
for all $\psi$ in the domain of $\phi(f)$; the subspace $\mathcal{D}$
of $\Hc$ generated by $\Omega$ and all $\phi(f_1) \cdots \phi(f_n)\Omega$
(the ``Wightman domain'')
turns out to be an invariant common domain of essential selfadjointness for 
all $\phi(f)$ \cite{ArakiYamagami:1982}. Furthermore, the
$\phi(f)$ are affiliated to the local von Neumann algebras 
$\Nc(O)$ in the GNS representation if ${\rm supp}(f) \subset O$ since their unitary groups
${\rm e}^{it \phi(f)} = \pi({\sf W}( t [f] ))$, $t \in \mathbb{R}$, are clearly contained
in $\Nc(O)$ for real-valued test-functions $f$ supported in $O$.

It is then easy to check that the quantum field thus obtained in the GNS representation of $\omega$
fulfills the following conditions:
\begin{itemize}
\item[(a)] $\phi(f)$ is complex linear in $f$ (by complex linear extension of $\phi(f)$ for real $f$)
\item[(b)] $\phi(f)^* = \phi(\bar{f})$ 
\item[(c)] $[\phi(f),\phi(h)] = i\sigma([f],[h]) {\bf 1}$
\item[(d)] $\phi((\Box + m^2)f) = 0$
\end{itemize}
where these relations hold for all $f,h \in C_0^\infty(M)$ as operator identities on $\mathcal{D}$.
Moreover, it holds that $W(f,h) = \ip{ \Omega}{ \phi(f)\phi(h) \Omega}$.

A slightly different formulation of the quantized Klein-Gordon
field  was used in \cite{CF-RV-ultraSJ} and, implicitly, also in Sections~\ref{sect:fluctuations} 
and~\ref{sect:further_SJ} of the present paper. There, we considered the abstract unital $*$-algebra 
$\mathscr{F}(M)$ generated by elements $\phi(f)$ ($f\in C_0^\infty(M)$) subject to relations (a)--(d) above.
Evidently, the Hilbert space $\mathcal{H}$ constructed above carries a representation 
of $\mathscr{F}(M)$ in an obvious way, and $\Omega$ determines a quasifree state on $\mathscr{F}(M)$ with two-point function $W$.
We also see that, in this sense, the algebra of fields obtained from the Weyl algebra is independent of the 
quasifree state used in the construction. Moreover, for quasi-free states, one may use
the Weyl and field algebras almost interchangeably.

\section{Decay estimate for the smeared field}\label{appx:decay}

We show, subject to the hypotheses of Theorem~\ref{thm:equiv1}, that the function
\[
\Upsilon(\mu) = 
\sum_{j\in J} \frac{1}{2\omega_j}
\left|\alpha_j \hat{f}(\mu+\omega_j) +\overline{\beta_j}\hat{f}(\mu-\omega_j)\right|^2
\] 
decays rapidly as  $\mu\to+\infty$, for all $M\in\NN$.

Let $N\in\NN$ be chosen so that  
$C:=\sum_{j\in J} (\omega_j^2+m^2)^{-2N}<\infty$. 
As $\alpha_j\to 1$, and $f$ is smooth and compactly supported there is, 
for each $M\in \NN$, a constant $C_M$ so that
\[
|\alpha_j \hat{f}(\mu+\omega_j)| \le \frac{C_M}{((\mu+\omega_j)^2+m^2)^{2N+M}} \le
\frac{C_M}{(\omega_j^2+m^2)^{2N}\mu^{2M}}
\]
for any $\mu>0$. Thus we have
\[
\sum_{j\in J} \frac{1}{2\omega_j}
\left|\alpha_j \hat{f}(\mu+\omega_j)\right|^2 = O(\mu^{-4M}), 
\qquad
\sum_{j\in J} \frac{1}{2\omega_j}
\left|\alpha_j \hat{f}(\mu+\omega_j)\right| = O(\mu^{-2M})
\]
As $\sup_{j\in J}|\beta_j \hat{f}(\mu-\omega_j)|\le
\sup_{j\in J}|\beta_j| \sup_{\omega} |\hat{f}(\omega)|<\infty$, 
it follows  that 
\begin{equation}\label{eq:part1}
\Upsilon(\mu) \le
\sum_{j\in J} \frac{1}{2\omega_j}
\left|\overline{\beta_j}\hat{f}(\mu-\omega_j)\right|^2
+O(\mu^{-2M}) \qquad (\mu\to +\infty).
\end{equation}

Now because of the rapid decrease of the $\beta_j$ and $\hat{f}$, there
are positive constants $C'_M$ and $C''_M$ such that 
\begin{align*}
(\omega_j^2+m^2)^{N}(\omega_j+m)^M|\beta_j| &\le C'_M \qquad (j\in J)\\
(m^2+\omega^2)^M|\hat{f}(\omega)| &\le C''_M \qquad (\omega\in\RR).
\end{align*}
Hence
\begin{align}\label{eq:beta_bound}
\sum_{j\in J} \frac{1}{2\omega_j}
\left|\overline{\beta_j}\hat{f}(\mu-\omega_j)\right|^2 &\le
\frac{(C'_M C''_M)^2}{2m} \sum_{j\in J} \frac{1}{(\omega_j^2+m^2)^{2N}} \frac{1}{(\omega_j+m)^{2M}((\omega_j-\mu)^2+m^2)^{2M}}
\notag \\
&\le\frac{(C'_M C''_M)^2 C}{2m} \sup_{j\in J}  \frac{1}{(\omega_j+m)^{2M}((\omega_j-\mu)^2+m^2)^{2M}} \notag \\
&\le\frac{ 2^{2M-1}(C'_M C''_M)^2 C}{\mu^{2M}m^{4M+1}}
\end{align}
for all $\mu>m$. 
In the last step, we used the following elementary observations. Let $K_{x_0}(x)=(1+x)(1+(x-x_0)^2)$, 
for $x_0>1$. Then, on $[0,\infty)$, $K_{x_0}$ is easily seen to have a local maximum at $x=x_{-}$ and global minimum at $x_{+}$,
where
\[
x_\pm = \frac{1}{3}\left(2x_0 - 1 \pm \sqrt{(x_0+1)^2-3}\right).
\]
Then $K_{x_0}(x_+)=(1+x_+)(1+(x-x_+)^2)\ge 1+x_+ >2x_0/3$,
so, in particular, $\inf_{\RR^+} K_{x_0}>x_0/2$. 
Moreover, 
as $K_{x_0}$ is strictly positive on $\RR^+$, the same classification of critical
points applies to positive integer powers of $K_{x_0}$. In particular, 
the minimum value of $K_{x_0}(x)^{2M}$ is $K_{x_0}(x_+)^{2M} > (x_0/2)^{2M}$.
This estimate is precisely what is needed to obtain~\eqref{eq:beta_bound}. In conjunction with \eqref{eq:part1},
we have shown that $\Upsilon(\mu)=
O(\mu^{-2M})$ as $\mu\to +\infty$, for all $M\in\NN$, as desired.

\end{document}